\documentclass[10pt,letterpaper]{article}
    \usepackage[top=0.85in,left=2.75in,footskip=0.75in]{geometry}

    \usepackage{pdfpages}

    \usepackage{amsmath,amssymb}

    \usepackage{changepage}

    \usepackage[utf8x]{inputenc}

    \usepackage{textcomp,marvosym}

    \usepackage{cite}

    \usepackage{nameref,hyperref}

    \usepackage[right]{lineno}

    \usepackage{microtype}
    \DisableLigatures[f]{encoding = *, family = * }

    \usepackage{array}

    \newcolumntype{+}{!{\vrule width 2pt}}

    \newlength\savedwidth

    \raggedright
    \usepackage[aboveskip=1pt,labelfont=bf,labelsep=period,justification=raggedright,singlelinecheck=off]{caption}

    \makeatletter
    \renewcommand{\@biblabel}[1]{\quad#1.}
    \makeatother

    \usepackage{lastpage,fancyhdr,graphicx}
    \usepackage{epstopdf}
    \fancyheadoffset[L]{2.25in}
    \fancyfootoffset[L]{2.25in}
    \usepackage{blkarray}
    \usepackage{algorithm}
    \usepackage{algorithmic}
    \usepackage{multirow}
    \usepackage{subfig}

\begin{document}
    \vspace*{0.2in}

    \begin{flushleft}
    {\Large
    \textbf\newline{Close spatial arrangement of mutants favors and disfavors fixation} 
    }
    \newline
    \\
    Yunming Xiao,
    Bin Wu\textsuperscript{*}
    \\
    \bigskip
    School of Sciences, Beijing University of Posts and Telecommunications, China
    \\
    \bigskip

    %
    %





    * bin.wu@bupt.edu.cn

    \end{flushleft}
    \section*{Abstract}
    Cooperation is ubiquitous across all levels of biological systems ranging from microbial communities to human societies.
    It, however, seemingly contradicts the evolutionary theory,
    since cooperators are exploited by free-riders and thus are disfavored by natural selection.
    Many studies based on evolutionary game theory have tried to solve the puzzle and figure out the reason why cooperation exists and how it emerges.
    Network reciprocity is one of the mechanisms to promote cooperation,
    where nodes refer to individuals and links refer to social relationships.
    The spatial arrangement of mutant individuals, which refers to the clustering of mutants, plays a key role in network reciprocity.
    Besides, many other mechanisms supporting cooperation suggest that the clustering of mutants plays an important role in the expansion of mutants.
    However, the clustering of mutants and the game dynamics are typically coupled.
    It is still unclear how the clustering of mutants alone alters the evolutionary dynamics.
    To this end, we employ a minimal model with frequency independent fitness on a circle.
    It disentangles the clustering of mutants from game dynamics. 
    The distance between two mutants on the circle is adopted as a natural indicator for the clustering of mutants or assortment. 
    We find that the assortment is an amplifier of the selection for the connected mutants compared with the separated ones. 
    Nevertheless, as mutants are separated, the more dispersed mutants are, the greater the chance of invasion is.
    It gives rise to the non-monotonic effect of clustering, which is counterintuitive.
    On the other hand, we find that less assortative mutants speed up fixation.
    Our model shows that the clustering of mutants plays a non-trivial role in fixation,
    which has emerged even if the game interaction is absent.

    \section*{Author summary}
    Evolutionary dynamics on networks are key for biological and social evolution.
    Typically, the clustering mutants on networks can dramatically alter the direction of selection.
    Previous studies on the assortment of mutants assume that individuals interact in a frequency-dependent way.
    It is hard to tell how assortment alone alters the evolutionary fate.
    We establish a minimal network model to disentangle the assortment from the game interaction.
    We find that for weak selection limit, the assortment of mutants plays little role in fixation probability.
    For strong selection limit, connected mutants, i.e., the maximum assortment, are best for fixation.
    When the mutants are separated by only one wild-type individual,
    it is worse off than that separated by more than one wild-type individual in fixation probability.
    Our results show the nontrivial yet fundamental effect of the clustering on fixation.
    Noteworthily, it has already arisen, even if the game interaction is absent.

    \section*{Introduction}
    Cooperation is ubiquitous in the natural world ranging from microbial communities to human societies.
    Yet, it is seemingly against evolutionary theory,
    since cooperators forgo their own interest to benefit others whereas defectors pay nothing to get the benefit.
    The past two decades have seen an intensive study on how cooperation evolves via natural selection \cite{five_rules, hauert2009, rick2001nature, proc_06_circle, kamran2015royal, Hamilton1964jtb, langer2008jtb, assort2012, hauert2004nature, lega2003evolution}.
    One of the key mechanisms to promote cooperation is network reciprocity.
    It assumes that individuals only interact with their neighbors.
    Consequently, either reproduction or competition for survival happens locally,
    which is not true for evolutionary dynamics in well-mixed population \cite{five_rules, proc_06_circle, langer2008jtb, hauert2004nature, lega2003evolution}.

    For network reciprocity, a simple rule has been derived \cite{rule} that
    cooperation is favored provided the benefit-to-cost ratio exceeds the average number of neighbors per individual.
    It holds for the Death-birth (DB) process under weak selection limit.
    A key intermediate step to achieve this simple rule is that
    a cooperator has more cooperator neighbors than defector neighbors.
    Furthermore, the fewer neighbors a cooperator has, the more proportion of cooperator neighbors a cooperator has.
    In other words, few neighbors per individual lead to the clustering of the cooperators for evolutionary dynamics on a network.
    A cooperator surrounded by many cooperator neighbors obtains high payoff,
    which facilitates the fixation of cooperation.
    This simple rule also paves the way to solve social dilemmas including those modeled by multi-player games \cite{wu2016plos}.
    Therefore, the assortment of cooperators has been intensively employed to investigate the fixation probability and the fixation time for stochastic evolutionary game dynamics on a network \cite{assort1, assort2, assort2012, wu2015pre, hauert2004nature, lega2003evolution}.
    Besides, other mechanisms promoting cooperation also result in the assortment of cooperators  as a key intermediate step \cite{rick2001nature, tag2, tag3},
    which is similar to the network reciprocity.
    Therefore, it would be necessary to investigate how the assortment alters the evolutionary outcome.

    For previous studies on the evolution of cooperation on a network \cite{proc_06_circle, langer2008jtb, hauert2004nature, lega2003evolution},
    both the game interaction and assortment are taken into account.
    Typically cooperation is modeled as a social dilemma via dyadic or multi-player games \cite{nowak2006book}.
    The assortment of cooperators follows as a result of evolutionary dynamics (for an exception, see \cite{bin2018aspiration}).
    It is still far from clear how assortment alone changes the fate of evolution.
    Here, we disentangle the game dynamics and the spatial clustering.
    And we establish a minimal model to explore this issue.
    To this end,
    we only consider the frequency-independent cases, without any game interactions, to explore the role that the clustering plays alone.
    As a first step, we adopt a circle as the underlying population structure.
    Our study starts with two mutants.
    They have an initial distance denoting the number of wild-type individuals between them.
    We regard the distance as a measure of the spatial assortment. 
    And we explore how the assortment of mutants alters the fixation probability and fixation time analytically.

    \section*{Models}
    \subsection*{Connected Mutants}
    \label{db}
\begin{figure}[!ht]
        \begin{adjustwidth}{-2.25in}{0in}
        \centering
        \includegraphics[width=1\linewidth]{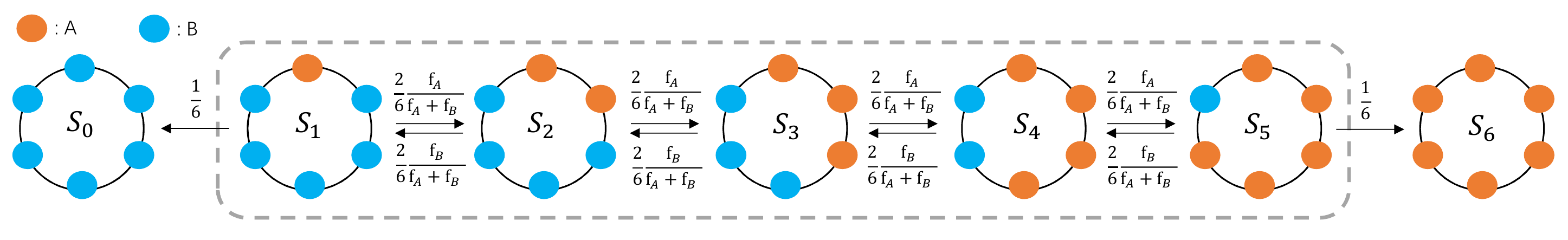}
        \caption{{\bf Markov chain for the Death-birth process.}
        The states in the dashed box belong to the same equivalence class of the Markov chain, whereas two states outside the box are absorbing states, respectively.
        The Markov chain is one-dimensional,
        thus the fixation probability starting from an arbitrary state can be analytically solved.
        Here, the population size is six, i.e., $N=6$.
        }
        \label{Figure1}
        \end{adjustwidth}
    \end{figure}

    We assume that there are $N$ individuals with two strategies, $A$ (wild-type) and $B$ (mutant). 
    The corresponding fitnesses are $f_{A}$ and $f_{B}$, respectively. 
    The fitness is frequency-independent.
    In other words, it is solely determined by the focal individuals's strategy,
    and has nothing to do with its neighbors'.
    All the individuals are located on a ring, i.e., every individual has exactly two neighbors. We consider the Death-birth (DB) process. For each round, an individual is randomly chosen to die. Its two neighbors compete to reproduce an offspring who adopts the same strategy as its parent. The chance of successful reproduction is proportional to the neighbors' fitnesses  (see Fig. \ref{Figure1} for illustration). 
    $w$ denotes the number of mutants, and $S_{w}$ is a state.
    Then the DB process is described by a one-dimensional Markov chain.
    The Markov chain has two absorbing states ($S_{0}$ and $S_{6}$) and the other states ($S_{i}$, where $1 \leq i \leq 5$) are of one equivalence class.

    Denote $P_{a, b}$ as the transition probability from state $S_{a}$ to $S_{b}$, the Kolmogorov backward equation is written as
    \begin{equation}
        \label{CK_one_dimensional}
        \begin{aligned}
        & \pi_{w} = P_{w, w+1}\pi_{w+1} + P_{w, w-1}\pi_{w-1} + (1 - P_{w, w+1} - P_{w, w-1})\pi_{w}, \\
            & \text{with}\ \pi_{0} = 0\ \text{and}\ \pi_{N} = 1,
        \end{aligned}
    \end{equation}
    where $\pi_{w}$ is the fixation probability starting from $S_{w}$. The fixation probability is then obtained \cite{firstcourse1975}:
    \begin{equation}
      \label{continuous_fixation}
        \pi_{w} = \frac{1 + \sum_{j=1}^{w-1}\prod_{k=1}^{j}\gamma_{k}}{1 + \sum_{w=1}^{N-1}\prod_{k=1}^{j}\gamma_{k}},
    \end{equation}
    where
    \begin{equation}
        \gamma_{w} = \frac{P_{w, w-1}}{P_{w, w+1}}.
    \end{equation}
    Let $r$ be the ratio of fitnesses between wild-type and mutant, i.e., $\frac{f_{A}}{f_{B}} = r$. It holds as follows:
    \begin{equation}
      \label{continuous_fixation_gamma}
      \gamma_{w} = \left\{
          \begin{aligned}
              & \frac{r+1}{2r}, & w = 1  \\
              & \frac{1}{r}, & w = 2, \ldots, N-2 \\
              & \frac{2}{r+1}, & w = N - 1
          \end{aligned}
          \right..
    \end{equation}
    Taking Eq. \eqref{continuous_fixation_gamma} into Eq. \eqref{continuous_fixation} leads to the fixation probabilities.
    For the population size $N = 6$, the fixation probability for two connected mutants is given:
    \begin{equation}
        \label{fp_6_0}
        \pi_{2} = \frac{r^{3}(1+3r)}{3+2r+2r^{2}+2r^{3}+3r^{4}}.
    \end{equation}

    Let $\tau_{i}^{A}$ denote the conditional fixation time from state $S_{i}$ to $S_{N}$, which refers to the mean time to absorption in state $S_{N}$ given the process starts in state $S_{i}$ and eventually reaches state $S_{N}$. We have
    \begin{equation}
        \label{conditional_fixation_time_eq}
        \begin{aligned}
        & \pi_{i}\tau_{i}^{A} = P_{i, i-1}\pi_{i-1}(\tau_{i-1}^{A} + 1) + (1 - P_{i,i-1} - P_{i,i+1})\pi_{i}(\tau_{i}^{A} + 1) +  P_{i,i+1}\pi_{i+1}(\tau_{i+1}^{A} + 1), \\
            & \text{with}\ \pi_{0}\tau_{0}^{A} = 0\ \text{and}\  \tau_{N}^{A} = 0.
        \end{aligned}
    \end{equation}
    Let us denote $\theta_{i}=\pi_{i} \tau_{i}^{A}$, then we arrive at a difference equation
    $\theta_{i} = P_{i, i-1}\theta_{i-1} + (1 - P_{i,i-1} - P_{i,i+1})\theta_{i}+  P_{i,i+1}\theta_{i+1}+1$ with boundary conditions $\theta_0 = 0$  and $\theta_N = 0$ \cite{hauert2009}.
    In particular, for $\theta_0=\pi_{0}\tau_{0}^{A}$,  $\tau_{0}^{A}$ is infinitely large since it takes forever for the mutant to fixate if there is no mutant initially. On the other hand, $\pi_0=0$. We thus assume $\theta_0=0$ as in \cite{hauert2009}.  
    Solving the recursive equations \cite{hinderson2014counter,hauert2009} leads to
    \begin{equation}
        \begin{aligned}
        & \tau_{i}^{A} = \tau_{1}^{A} \frac{\pi_{1}}{\pi_{i}} \sum_{k=1}^{i-1}\prod_{m=1}^{k-1}\gamma_{m} - \sum_{k=1}^{i-1}\sum_{l=1}^{k-1}\frac{1}{\pi_{i}}\frac{\pi_l}{P_{l,l+1}}\prod_{m=l+1}^{k}\gamma_m, \\
            & \text{with}\ \tau_{1} = \sum_{k=1}^{N-1}\sum_{l=1}^{k}\frac{\pi_l}{P_{l,l+1}}\prod_{m=l+1}^{k}\gamma_m.
        \end{aligned}
    \end{equation}
    Taking $N=6$ into the above equation, we obtain
    \begin{equation}
        \label{ft_6_0}
        \tau_{2}^{A} = \frac{3(11+75r+132r^{2}+140r^{3}+109r^{4}+45r^{5})}{(1+3r)(3+2r+2r^{2}+2r^{3}+3r^{4})}.
    \end{equation}

    \subsection*{Separated Mutants For Small Circle}
    \label{db_seperate}
\begin{figure}[!ht]
        \begin{adjustwidth}{-2.25in}{0in}
        \centering
        \includegraphics[width=\linewidth]{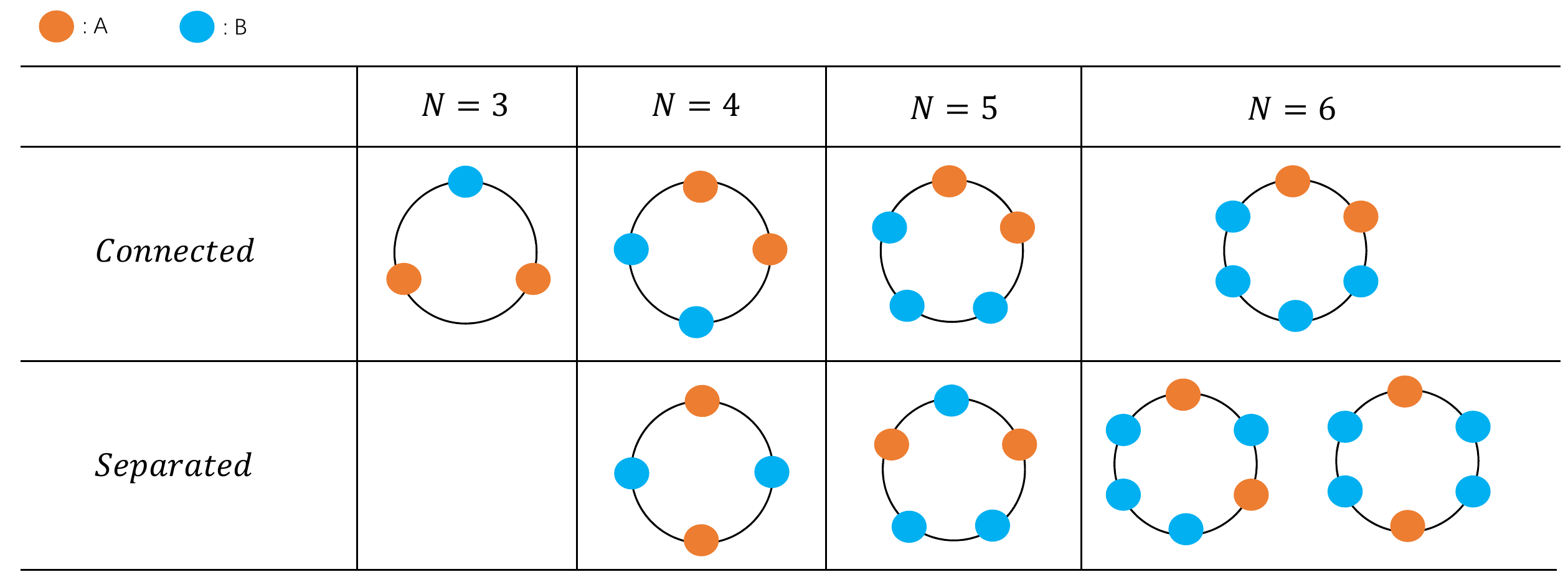}
        \caption{{\bf Network configuration for two mutants.}
        If the population size is three, the two mutants have to be connected.
        If the population size is four or five,
        the two mutants can be separated by at most one wild-type individual.
        If the population size is six,
        the two mutants can be of distance zero, one and two,
        i.e., three types.
        In other words, six is the minimum population size of a circle,
        which gives rise to three distances between two mutants.
        Thus we adopt the population size six as an illustration model.
        }
        \label{Figure2}
        \end{adjustwidth}
    \end{figure}

    To explore the effect of the spatial clustering, we consider the process that there are two mutants with distance $d$ in the beginning. That is to say, there are $d$ connected wild-type individuals located between two mutants initially. In this section, we take $N=6$ as an illustrative case. Note that six is the minimal size of a circle, in which there are two kinds of unconnected mutants. All the circles with population size below six have none or one such network configurations, as shown in Fig. \ref{Figure2}.

\begin{figure}[!ht]
        \begin{adjustwidth}{-2.25in}{0in}
        \centering
        \includegraphics[width=\linewidth]{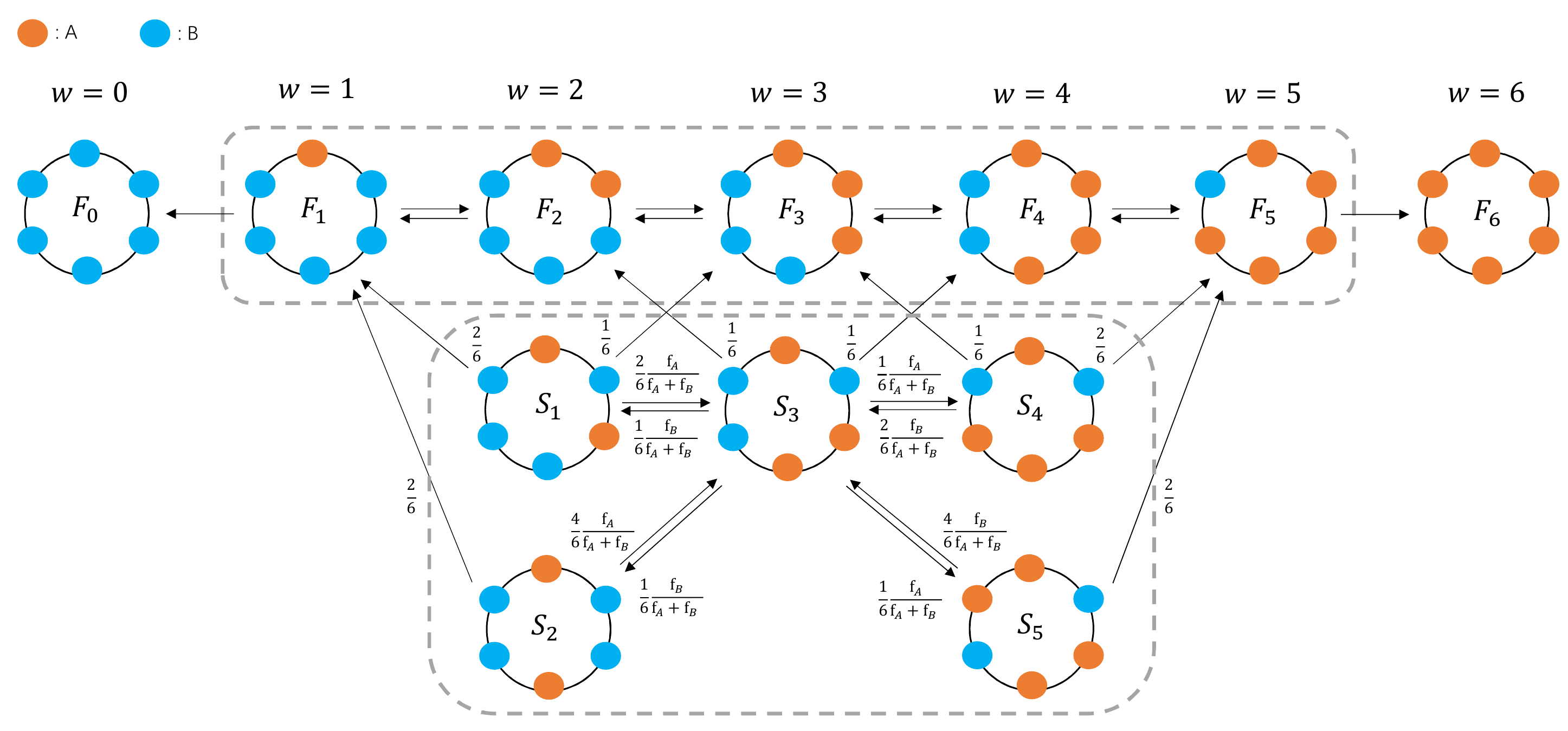}
        \caption{{\bf The Markov chain of the Death-birth process with separated mutants.}
        In contrast with the Markov chain with connected mutants alone,
        the underlying Markov chain is no longer one-dimensional.
        The transient states are further categorized into two equivalence classes, denoted as $S$ and $F$, respectively.
        Each one is grouped by the dashed boxes.
        The $S$ class will sooner or later enter the $F$ class,
        whereas the $F$ class cannot go to the $S$ class. 
        Instead, they will enter the absorbing states sooner or later.
        $w$ refers to the number of mutants.
        }
        \label{Figure3}
        \end{adjustwidth}
    \end{figure}

    As illustrated in Fig. \ref{Figure3}, the process gives rise to more states than that which starts with two connected mutants. Comparing with the previous process in Fig. \ref{Figure1}, we divide all the states into two sets: the middle-state set $S$ and the final-state set $F$. The middle states refer to all the states with two separated groups of mutants whereas the final states contain only one mutant group. Note that a group refers to connected individuals with the same strategy. Fig. \ref{Figure3} shows four properties of the process:
    i) All the middle states reach each other and belong to one equivalence class.
    ii) The final states reach each other and belong to one equivalence class ($F_{i}, 1 \leq i \leq 5$) and two absorbing states (${F_{0}}$ and ${F_{6}}$).
    iii) The middle states reach final states in finite time; however, the final states cannot reach any middle states.
    iv) The middle states are transient (sooner or later, they walk into one of the final states).
    These four features imply that the underlying Markov chain is not one-dimensional anymore,
    which leads to both computational and analytical challenges.

    The transition matrix $P$ for the process in Fig. \ref{Figure3} is listed as follows:
    \begin{equation}
        \label{transfer_matrix_6}
        \scriptsize
        \begin{blockarray}{cccccc|ccccccc}
            & S_{1} & S_{2} & S_{3} & S_{4} & S_{5} & F_{0} & F_{1} & F_{2} & F_{3} & F_{4} & F_{5} & F_{6} \\
            \begin{block}{c(ccccc|ccccccc)}
                S_{1} & \frac{1}{6} \frac{3+r}{1+r} & 0 & \frac{1}{6}\frac{2r}{1+r} & 0 & 0 & 0 & \frac{1}{3} & 0 & \frac{1}{6} & 0 & 0 & 0 \\
                S_{2} & 0 & \frac{1}{6}\frac{4}{1+r} & \frac{1}{6}\frac{4r}{1+r} & 0 & 0 & 0 & \frac{1}{3} & 0 & 0 & 0 & 0 & 0 \\
                S_{3} & \frac{1}{6}\frac{1}{1+r} & \frac{1}{6}\frac{1}{1+r} & \frac{1}{6}\frac{2+2r}{1+r} & \frac{1}{6}\frac{r}{1+r} & \frac{1}{6}\frac{r}{1+r} & 0 & 0 & \frac{1}{6} & 0 & \frac{1}{6} & 0 & 0  \\
                S_{4} & 0 & 0 & \frac{1}{6}\frac{2}{1+r} & \frac{1}{6}\frac{2+r}{1+r} & 0 & 0 & 0 & 0 & \frac{1}{6} & 0 & \frac{1}{3} & 0 \\
                S_{5} & 0 & 0 & \frac{1}{6}\frac{4}{1+r} & 0 & \frac{1}{6}\frac{4r}{1+r} & 0 & 0 & 0 & 0 & 0 & \frac{1}{3} & 0 \\ \cline{1-13}
                F_{0} & 0 & 0 & 0 & 0 & 0 & 1 & 0 & 0 & 0 & 0 & 0 & 0 \\
                F_{1} & 0 & 0 & 0 & 0 & 0 & \frac{1}{6} & \frac{1}{6}\frac{5+3r}{1+r} & \frac{1}{6}\frac{2r}{1+r} & 0 & 0 & 0 & 0 \\
                F_{2} & 0 & 0 & 0 & 0 & 0 & 0 & \frac{1}{6}\frac{2}{1+r} & \frac{4}{6} & \frac{1}{6}\frac{2r}{1+r} & 0 & 0 & 0 \\
                F_{3} & 0 & 0 & 0 & 0 & 0 & 0 & 0 & \frac{1}{6}\frac{2}{1+r} & \frac{4}{6} & \frac{1}{6}\frac{2r}{1+r} & 0 & 0 \\
                F_{4} & 0 & 0 & 0 & 0 & 0 & 0 & 0 & 0 & \frac{2}{6}\frac{1}{1+r} & \frac{4}{6} & \frac{2}{6}\frac{r}{1+r} & 0 \\
                F_{5} & 0 & 0 & 0 & 0 & 0 & 0 & 0 & 0 & 0 & \frac{1}{6}\frac{2}{1+r} & \frac{1}{6}\frac{3+5r}{1+r} & \frac{1}{6} \\
                F_{6} & 0 & 0 & 0 & 0 & 0 & 0 & 0 & 0 & 0 & 0 & 0 & 1 \\
            \end{block}
        \end{blockarray}
    \end{equation}

    Denote $\Psi_{i}$ as the fixation probability starting from state $i \in \{S, F\}$ and ending up with state $F_{6}$. Based on the Markov property, the following holds:
    \begin{equation}
        \Psi_{i} = \sum_{j \in \{S, F\}} P_{i, j} \Psi_{j}, \ \forall i \in \{S, F\}.
    \end{equation}
    It is equivalent to
    \begin{equation}
        \label{target}
        \Psi = P\Psi,
    \end{equation}
    with boundary conditions $\Psi_{F_{0}} = 0$ and $\Psi_{F_{6}} = 1$ (subject to the property ii)).

    We divide the states into two sets: the middle-state set $S$ and the final-state set $F$. We denote
    $\Psi = \left(\begin{matrix}
        \Psi_{S}\\
        \Psi_{F}
    \end{matrix}\right)$.
    As the two crossing lines in Eq. \eqref{transfer_matrix_6} illustrates, the one-step transition matrix $P$ can be written as  
    \begin{equation}
        \label{transfer_matrix}
        P =
        \begin{blockarray}{ccc}
            & S & F \\
            \begin{block}{c(cc)}
                S & Q_{1} & Q_{2} \\
                F & 0 & Q_{3} \\
            \end{block}
        \end{blockarray}
        .
    \end{equation}
    The transition probability from $F$ to $S$ is a zero matrix,
    which arises from property iii).
    In addition, we have that the sub-matrix $Q_{2}$ is independent of mutant fitness $r$.
    In fact, the entries in $Q_2$ implies the transition probability whose event is the collapse of two separate groups with the same strategy. Here, a group refers to connected individuals with the same strategy. 
    Take the transition from $S_{1}$ to $F_{1}$ as an example, the transition occurs when a mutant is chosen to die with probability $\frac{2}{6}$. The chosen mutant has two wild-type neighbors. In this case, the chosen mutant will, with probability one, be replaced by a wild-type offspring. Thus, the transition probability from $S_{1}$ to $F_{1}$ is independent on the relative fitness of the mutant $r$. In general, this applies to any transition from the middle-state set $S$ to final-state set $F$. Therefore, $Q_{2}$ is independent of mutant fitness $r$. Similarly, $Q_{1}$ and $\Psi$ are dependent on $r$.

    Taking Eq. \eqref{transfer_matrix} into Eq. \eqref{target}, we obtain
    \begin{equation}
        \left\{
            \begin{aligned}
                \Psi_{S} &= Q_{1}\Psi_{S} + Q_{2}\Psi_{F} \\
                \Psi_{F} &= Q_{3}\Psi_{F}
            \end{aligned}
        \right..
    \end{equation}
    Note that the second equation ($\Psi_{F} = Q_{3}\Psi_{F}$) is the same as Eq. \eqref{CK_one_dimensional}. Thus we have $\Psi_{F} = \pi$.

    We now consider the first equation ($\Psi_{S} = Q_{1}\Psi_{S} + Q_{2}\Psi_{F}$), which can be transferred to $(I - Q_{1})\Psi_{S} = Q_{2}\Psi_{F}$. We show that $I - Q_{1}$ is invertible in the following. Since all the middle states are transient with respect to process property i) and iv), we have $\sum_{t=0}^{\infty}P_{ij}^{(t)} < \infty,\ \forall i, j \in S$ \cite{grimmett2009book}.
    This is equivalent to
    \begin{equation}
        \left( \sum_{t=0}^{\infty}Q_{1}^{t} \right)_{ij} < \infty.
    \end{equation}
    Let $H = \sum_{t=0}^{\infty}Q_{1}^{t}$, and we know $H$ exists. Notice that
    \begin{equation}
        H(I - Q_{1}) = \sum_{t=0}^{\infty}Q_{1}^{t} (I - Q_{1}) = \sum_{t=0}^{\infty}Q_{1}^{t} - \sum_{t=1}^{\infty}Q_{1}^{t} = I,
    \end{equation}
    where $I$ is the identity matrix with the same size as that of $Q_{1}$. This shows that $H$ is the left-inverse of $(I - Q_{1})$, and a similar argument shows that $H$ is also the right-inverse of $(I - Q_{1})$. We then acknowledge that $H = (I-Q_{1})^{-1}$. Thus, $(I-Q_{1})$ is invertible.

    Thus, it holds
    \begin{equation}
        \label{target_psi}
        \Psi_{S} = (I - Q_{1})^{-1} Q_{2} \Psi_{F}.
    \end{equation}
    In the process of Fig. \ref{Figure3}, the fixation probabilities of states with two separated mutants are listed as follows:
    \begin{equation}
        \label{specific_case}
        \left\{
        \begin{aligned}
            & \Psi_{S_{1}} = \frac{r^{2}(28 + 167 r + 475 r^{2} + 920 r^{3} + 1036 r^{4} + 585 r^{5} + 117 r^{6})}{(3+2r+2r^{2}+2r^{3}+3r^4)(45+210r+322r^{2}+210r^{3}+45r^4)} \\
            & \Psi_{S_{2}} = \frac{r^{2}(39 + 179 r + 456 r^{2} + 896 r^{3} + 1041 r^{4} + 597 r^{5} + 120 r^{6})}{(3+2r+2r^{2}+2r^{3}+3r^4)(45+210r+322r^{2}+210r^{3}+45r^4)}
        \end{aligned}
        \right..
    \end{equation}

    We now investigate how long it takes for mutants to reach the state consisting of only mutants.
    Let $\tau_{i}^{A}$ be the average conditional fixation time from state $i \in \{S, F\}$ to $F_{6}$, given the population ends up with all mutants, i.e, $F_{6}$. $T^{A}$ is a vector of $\tau_{i}^{A}$, and it is denoted as $T^{A} = \left(\begin{matrix}
        T_{S}^{A}\\
        T_{F}^{A}
    \end{matrix}\right)$, where $T_{S}^{A}$ is the conditional fixation time to $F_{6}$ for the middle states and $T_{F}^{A}$ is that for the final states. For state $i \in \{S, F\}$, we have
    \begin{equation}
        \label{time1}
        \begin{aligned}
        & \Psi_{i} \cdot \tau_{i}^{A} = \sum_{j \in \{S, F\}} \Psi_{j} \cdot P_{i, j} \cdot (\tau_{j}^{A} + 1), \\
            & \text{with}\ \tau_{F_{6}}^{A} = 0\ \text{and}\ \Psi_{F_{0}} \cdot \tau_{F_{0}}^{A} = 0.
        \end{aligned}
    \end{equation}
    The right side of Eq. \eqref{time1} contains all the cases that one-step further from $S_{1}$, weighted by the one-step transition probabilities.

    The symbol $\circ$ is the Hadamard product. For two matrices $A = [a_{ij}]$ and $B = [b_{ij}]$ with the same dimensions, we have $A \circ B = [a_{ij} \cdot b_{ij}]$. We then transfer Eq. \eqref{time1} to
    \begin{equation}
        \Psi \circ T^{A} = P \cdot [ \Psi \circ (T^{A} + \textbf{1}) ],
    \end{equation}
    where $\textbf{1}$ is a vector of value $1$ with the same dimensions as $T^{A}$. This is equivalent to
    \begin{equation}
        \Psi \circ T^{A} = P \cdot (\Psi \circ T^{A}) + P \cdot \Psi.
    \end{equation}
    And moving all elements with $\Psi \circ T$ to the left side, we have
    \begin{equation}
        \label{time2}
        (I - P) \cdot (\Psi \circ T^{A}) = P \cdot \Psi,
    \end{equation}
    where $I$ is the identity matrix.

    Splitting the middle and final states, Eq. \eqref{time2} is written as 
    \begin{equation}
        \left( \begin{matrix} I - Q_{1} & -Q_{2} \\ 0 & I - Q_{3} \end{matrix} \right)
        \cdot
        \left( \begin{matrix} \Psi_{S} \circ T_{S}^{A} \\ \Psi_{F} \circ T_{F}^{A} \end{matrix} \right)
        =
        \left( \begin{matrix} Q_{1} & Q_{2} \\ 0 & Q_{3} \end{matrix} \right)
        \cdot
        \left( \begin{matrix} \Psi_{S} \\ \Psi_{F} \end{matrix} \right).
    \end{equation}
    We then obtain
    \begin{equation}
        \label{target1}
        \left\{
        \begin{aligned}
            & (I - Q_{1}) \cdot (\Psi_{S} \circ T_{S}^{A}) - Q_{2} \cdot (\Psi_{F} \circ T_{F}^{A}) = Q_{1} \cdot \Psi_{S} + Q_{2} \cdot \Psi_{F} \\
            & 0 - (I - Q_{3}) \cdot (\Psi_{F} \circ T_{F}^{A}) = 0 + Q_{3} \cdot \Psi_{F}
        \end{aligned}
        \right..
    \end{equation}
    We have a solution for Eq. \eqref{target1} based on \cite{hauert2009}. In particular, for $j = 1, \ldots, 5$, we have
    \begin{equation}
        \tau_{F_{j}}^{A} = \tau_{F_{1}}^{A} \frac{\Psi_{F_{1}}}{\Psi_{F_{j}}} \sum_{k=1}^{j-1} \prod_{m=1}^{k-1} \gamma_{m} - \sum_{k=1}^{j-1} \sum_{l=1}^{k-1} \frac{\Psi_{F_{l}}}{\Psi_{F_{j}}} \frac{1}{P_{F_{l}, F_{l+1}}} \prod_{m=l+1}^{k} \gamma_{m}.
    \end{equation}

    We now look into the first equation in Eq. \eqref{target1}. Note that we have proved $(I - Q_{1})$ is invertible, thus we have
    \begin{equation}
        \label{tsa}
        T_{S}^{A} = (I - Q_{1})^{-1} (Q_{1} \cdot \Psi_{S} + Q_{2} \cdot \Psi_{F} + Q_{2} \cdot (\Psi_{F} \circ T_{F}^{A})) \oslash \Psi_{S},
    \end{equation}
    where $\oslash$ is the Hadamard division operator. And Eq. \eqref{tsa} is equivalent to
    \begin{equation}
        \label{ft_25_1}
        T_{S}^{A} = (I - Q_{1})^{-1} [Q_{1} \cdot (I - Q_{1})^{-1}Q_{2}\Psi_{F} + Q_{2} \cdot \Psi_{F} + Q_{2} \cdot (\Psi_{F} \circ T_{F}^{A})] \oslash [(I - Q_{1})^{-1}Q_{2}\Psi_{F}].
    \end{equation}
    We do not present the analytic expressions here due to the great complexity of the expression of $\tau_{S_{1}}^{A}$ and $\tau_{S_{2}}^{A}$.

    \subsection*{Separated Mutants For Large Circles}
    \label{algorithm}
\begin{figure}[!ht]
        \centering
        \includegraphics[width=0.3\linewidth]{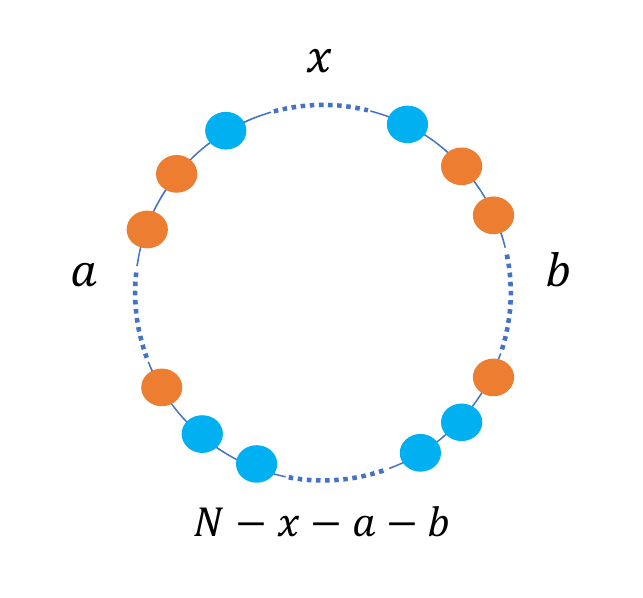}
        \caption{{\bf State notation of the configuration in a circle.}
        We assume that there are at most two separated mutant groups.
        The state is denoted as a triplet $(x,a,b)$.
        Here $x$ refers to the minimal distance between two mutant groups,
        $a$ and $b$ represent the group sizes of the smaller group and that of the larger one.
        The following inequalities holds: $a \leq b$ and $x \leq N-x-a-b$.
        }
        \label{Figure4}
    \end{figure}

    We now address the DB process on a circle for arbitrary size.
    We denote the population size as $N$. 
    A group refers to connected individuals with the same strategy.
    As Fig. \ref{Figure4} illustrated, each state corresponds to a triplet $(x, a, b)$: $x$ is the minimal distance of two mutant groups, $a$ is the population size of the smaller mutant group and $b$ is the population size of the larger mutant group. Note that the larger distance between two mutant groups equal to $N - x - a - b$.

    All the states for process for population size $N$ are listed in Table \ref{list_state}. We divide the states into the middle-state set $S$ and the final-state set $F$. When the minimal distance $x$ between two mutant groups is zero ($x = 0$), or when the number of one of the mutant groups is $0$ ($a = 0$), we use $F$ to replace the triplet expression $S_{x, a, b}$ ($S_{0, a, b} = F_{a+b}$ and $S_{x, 0, b} = F_{b}$). The middle states have two separate mutant groups whereas the final states only have one. From Table \ref{list_state}, we find that the total number of states is
    \begin{equation}
    \sum_{w=0}^{N}\lfloor\frac{w}{2}\rfloor\cdot\lfloor\frac{N-w}{2}\rfloor + N + 1.
    \end{equation}
    The total number of states is of $O(N^{2})$, since $(\sum_{w=0}^{N}\frac{w}{2}\cdot\frac{N-w}{2} + N + 1)$ is $O(N^{2})$. Difficulty arises to calculate the fixation probabilities with the equation $\Psi P = \Psi$.

    \begin{table}[!ht]
        \begin{adjustwidth}{-2.25in}{0in}
        \centering
        \caption{The states of the Markov chain in a circle population of size $N$. $w$ refers to the number of mutants.}
        \begin{tabular}{|c|c|c|c|cc|c|ccc|c|c|} \hline
        $w = 0$ & $1$ & $2$ & $3$ & \multicolumn{2}{c|}{$4$} & $\ldots$ & \multicolumn{3}{c|}{$w$} & $\ldots$ & $N$ \\\hline
        $F_{0}$ & $F_{1}$ & $F_{2}$ & $F_{3}$ & \multicolumn{2}{c|}{$F_{4}$} & $\ldots$ & \multicolumn{3}{c|}{$F_{w}$} & $\ldots$ & $F_{N}$ \\

        && $S_{1, 1, 1}$ & $S_{1, 1, 2}$ & $S_{1, 1, 3}$ & $S_{1, 2, 2}$ & $\ldots$ & $S_{1, 1, w-1}$  & $\ldots$ & $S_{1, \lfloor\frac{w}{2}\rfloor, w - \lfloor\frac{w}{2}\rfloor}$ & $\ldots$ & \\

        && $S_{2, 1, 1}$ & $S_{2, 1, 2}$ & $S_{2, 1, 3}$ & $S_{2, 2, 2}$ & $\ldots$ & $S_{2, 1, w-1}$  & $\ldots$ & $S_{2, \lfloor\frac{w}{2}\rfloor, w - \lfloor\frac{w}{2}\rfloor}$ & $\ldots$ & \\

        && $\ldots$ & $\ldots$ & \multicolumn{2}{c|}{$\ldots$} & $\ldots$ & \multicolumn{3}{c|}{$\ldots$} & $\ldots$ & \\

        && $S_{\lfloor\frac{N-2}{2}\rfloor, 1, 1}$
        & $S_{\lfloor\frac{N-3}{2}\rfloor, 1, 2}$
        & $S_{\lfloor\frac{N-4}{2}\rfloor, 1, 3}$
        & $S_{\lfloor\frac{N-4}{2}\rfloor, 2, 2}$
        & $\ldots$
        & $S_{\lfloor\frac{N-w}{2}\rfloor, 1, w-1}$
        & $\ldots$
        & $S_{\lfloor\frac{N-w}{2}\rfloor, \lfloor\frac{w}{2}\rfloor, w - \lfloor\frac{w}{2}\rfloor}$
        & $\ldots$ & \\\hline
        \end{tabular}
        \label{list_state}
        \end{adjustwidth}
    \end{table}

    For every transition between the states in Table \ref{list_state}, the state $S_{x, a, b}$ stays where it is, or transits to a state where the mutant number is one greater or one less. Note that the mutant number equals to the sum of two mutant group sizes $a + b$. Take $S_{1, 1, 1}$ for an example, it can transit to itself, $S_{1, 1, 2}$, $S_{2, 1, 2}$ or $F_{1}(=S_{1, 0, 1})$. The state transition only occurs when an individual at the border of a group is chosen to die. As the mutants and wild-type individuals have one or two groups respectively, there are at most $8$ individuals on the border. That is to say, starting from any state, there are at most $8$ transitions. If the population size $N$ is large, the transition matrix is sparse. We list all the transition probabilities and boundary conditions in Table \ref{transit_matrix}.

    \begin{table}[!ht]
        \begin{adjustwidth}{-2.25in}{0in}
        \centering
        \caption{
        Transition probabilities for the Markov chain starting from state $S_{x, a, b}$. 
        The first column indicates the state that $S_{x,a,b}$ transit to.
        The first row categorizes state $S_{x,a,b}$ by the index $(x,a,b)$.
        The transtion probabilities are shown in the rest of table.
        The population size is $N$. 
        }
        \begin{tabular}{|c|c|c|c|c|c|} \hline
        $P$ & $N-x-a-b=1$ & $x=1$ & $a=1$ & $b=1$ & $\text{Otherwise}$ \\\hline
        $F_{a+b+1}$ & $\frac{1}{N}$ & $\frac{1}{N}$ & $0$ & $0$ & $0$ \\\hline
        $F_{a}$ & $0$ & $0$ & $\frac{1}{N}$ & $0$ & $0$ \\\hline
        $F_{b}$ & $0$ & $0$ & $0$ & $\frac{1}{N}$ & $0$ \\\hline
        $S_{x,a+1,b}$ & $0$ & $\frac{1}{N}\frac{r}{1+r}$ & $\frac{1}{N}\frac{r}{1+r}$ & $\frac{1}{N}\frac{r}{1+r}$ & $\frac{1}{N}\frac{r}{1+r}$ \\\hline
        $S_{x,a,b+1}$ & $0$ & $\frac{1}{N}\frac{r}{1+r}$ & $\frac{1}{N}\frac{r}{1+r}$ & $\frac{1}{N}\frac{r}{1+r}$ & $\frac{1}{N}\frac{r}{1+r}$ \\\hline
        $S_{x-1,a+1,b}$ & $\frac{1}{N}\frac{r}{1+r}$ & $0$ & $\frac{1}{N}\frac{r}{1+r}$ & $\frac{1}{N}\frac{r}{1+r}$ & $\frac{1}{N}\frac{r}{1+r}$ \\\hline
        $S_{x-1,a,b+1}$ & $\frac{1}{N}\frac{r}{1+r}$ & $0$ & $\frac{1}{N}\frac{r}{1+r}$ & $\frac{1}{N}\frac{r}{1+r}$ & $\frac{1}{N}\frac{r}{1+r}$ \\\hline
        $S_{x,a-1,b}$ & $\frac{1}{N}\frac{1}{1+r}$ & $\frac{1}{N}\frac{1}{1+r}$ & $0$ & $\frac{1}{N}\frac{1}{1+r}$ & $\frac{1}{N}\frac{1}{1+r}$ \\\hline
        $S_{x+1,a-1,b}$ & $\frac{1}{N}\frac{1}{1+r}$ & $\frac{1}{N}\frac{1}{1+r}$ & $0$ & $\frac{1}{N}\frac{1}{1+r}$ & $\frac{1}{N}\frac{1}{1+r}$ \\\hline
        $S_{x,a,b-1}$ & $\frac{1}{N}\frac{1}{1+r}$ & $\frac{1}{N}\frac{1}{1+r}$ & $\frac{1}{N}\frac{1}{1+r}$ & $0$ & $\frac{1}{N}\frac{1}{1+r}$ \\\hline
        $S_{x+1,a,b-1}$ & $\frac{1}{N}\frac{1}{1+r}$ & $\frac{1}{N}\frac{1}{1+r}$ & $\frac{1}{N}\frac{1}{1+r}$ & $0$ & $\frac{1}{N}\frac{1}{1+r}$ \\\hline
        \end{tabular}
        \label{transit_matrix}
        \end{adjustwidth}
    \end{table}

    The process for arbitrary population size shares the same four properties as the process of population size $N=6$:
    i) The middle states reach any other middle states and belong to one equivalence class. Take the transition from $S_{1, 1, 1}$ to $S_{2, 2, 2}$ for an instance, there is a path as $S_{1, 1, 1} \to S_{1, 1, 2} \to S_{2, 1, 1} \to S_{2, 1, 2} \to S_{2, 2, 2}$. Going through this path, the transition number in Table \ref{transit_matrix} occurs in the order $\#5 \to \#11 \to \#5 \to \#4$.
    ii) The final states contain three equivalence classes. One is ${F_{0}}$, one is ${F_{N}}$, and all the rest give rise to the other equivalence class.
    iii) The middle states reach final states in finite time, whereas the final states cannot reach any middle states.
    iv) The middle states are transient states.

    With the four properties, the analysis from Eq. \eqref{target} to Eq. \eqref{target_psi} still apply here. We obtain the fixation probabilities of mutants for the process for arbitrary population size by
    \begin{equation}
        \label{target_psi_2}
        \Psi_{S} = (I - Q_{1})^{-1} Q_{2} \Psi_{F},
    \end{equation}
    where $Q_{1}$ and $\Psi_{F}$ are dependent on the relative mutant fitness $r$, whereas $Q_{2}$ is independent on $r$. Similarly, we obtain the conditional fixation time for mutants with arbitrary population size as
    \begin{equation}
        \label{target_t_2}
        T_{S}^{A} = (I - Q_{1})^{-1} [Q_{1} \cdot (I - Q_{1})^{-1}Q_{2}\Psi_{F} + Q_{2} \cdot \Psi_{F} + Q_{2} \cdot (\Psi_{F} \circ T_{F}^{A})] \oslash [(I - Q_{1})^{-1}Q_{2}\Psi_{F}].
    \end{equation}

    In \nameref{S2_Appendix}, we develop an algorithm to numerically obtain $Q_{1}$ and $Q_{2}$ with a time complexity of $O(N^{2})$ and a space complexity of $O(N^{4})$ ($O(N^{2})$ if sparse matrix method is employed). Combining with Eq. \eqref{target_psi_2} and Eq. \eqref{target_t_2}, we have the fixation probabilities and conditional fixation times for mutants with arbitrary population size $N$.
    As the algorithm makes use of matrix multiplications and inversions, the time complexities to obtain the fixation probability and conditional fixation time are both of $O(N^{4.746})$ \cite{algo_complex, algo_complex2, algo_complex3}.

    In particular, with Taylor's expansion around $r=1$ for Eq. \eqref{target_psi_2}, we have (see \nameref{S3_Appendix})
    \begin{equation}
        \Psi(r) = \Psi(1) + \frac{d}{d r}\Psi(r)\bigg| _{r=1}(r-1) + \frac{1}{2} \frac{d^{2}}{d r^{2}}\Psi(r)\bigg| _{r=1}(r-1)^{2} + o((r-1)^{2}),
    \end{equation}
    where
    \begin{small}
    \begin{align}
        \frac{d}{d r}\Psi(r) ={}& [I - Q_{1}(r)]^{-1}\frac{d}{d r}Q_{1}(r)[I - Q_{1}(r)]^{-1}Q_{2}\pi(r) + [I - Q_{1}(r)]^{-1}Q_{2}\frac{d}{d r}\pi(r),\label{1st_derivative}
        \\
        \begin{split}\label{2nd_derivative}
            \frac{d^{2}}{d r^{2}}\Psi(r)
            ={}& 2[I - Q_{1}(r)]^{-1}\frac{d}{d r}Q_{1}[I - Q_{1}(r)]^{-1}\frac{d}{d r}Q_{1}(r)[I - Q_{1}(r)]^{-1}Q_{2}\pi(r) \\
            &+ [I - Q_{1}(r)]^{-1}\frac{d^{2}}{d r^{2}}Q_{1}(r)[I - Q_{1}(r)]^{-1}Q_{2}\pi(r) \\
            &+ 2[I - Q_{1}(r)]^{-1}\frac{d}{d r}Q_{1}(r)[I - Q_{1}(r)]^{-1}Q_{2}\frac{d}{d r}\pi(r) \\
            &+ [I - Q_{1}(r)]^{-1}Q_{2}\frac{d^{2}}{d r^{2}}\pi(r).
        \end{split}
    \end{align}
    \end{small}
    The algorithm we developed also applies to calculate the derivatives of the fixation probabilities. 
    The derivative of a matrix is defined as the matrix of derivatives of corresponding item. 
    We obtain $\frac{d}{d r}Q_{1}(r)$ by turning values in Table \ref{transit_matrix} into their first-order derivatives and running through the algorithm in \nameref{S2_Appendix}. 
    Following Eq. \eqref{1st_derivative}, we obtain the first-order derivatives of the fixation probabilities. 
    Besides, the time complexity is the same order as that of the fixation probability. 
    Matrix multiplications and additions are required but they do not increase the time complexity. 
    The required space is doubled but it is still of complexity $O(N^4)$ ($O(N^{2})$ for adopting sparse matrices). 
    Similarly, we find that the higher-order derivatives of the fixation probabilities require only the same-order or lower-order derivatives of $Q_{1}$. 
    Thus, we obtain the second-order derivatives and higher-order ones by turning the values in Table \ref{list_state} to their higher-order derivatives. 
    The overall complexity stays the same.

    \section*{Results}
    \subsection*{Fixation Probabilities}
    We have already obtained the fixation probabilities of the mutants for a circle with population size $N=6$ based on Eq. \eqref{fp_6_0} and Eq. \eqref{specific_case}.
    Expanding the equations around neutral selection, i.e. $r=1$, gives rise to
    \begin{equation}
        \label{expand_6}
        \left\{
        \begin{aligned}
            & \Psi_{F_{2}} (= \pi_{2}) & = \frac{2}{6} + \frac{7}{12}(r-1) - \frac{40}{288}(r-1)^{2} + o((r-1)^{3}) \\
            & \Psi_{S_{1}} & = \frac{2}{6} + \frac{6}{12}(r-1) - \frac{49}{288}(r-1)^{2} + o((r-1)^{3}) \\
            & \Psi_{S_{2}} & = \frac{2}{6} + \frac{6}{12}(r-1) - \frac{46}{288}(r-1)^{2} + o((r-1)^{3})
        \end{aligned}
        \right..
    \end{equation}
    Note that $F_{2}$ refers to the state of two connected mutants, while $S_{1}$ and $S_{2}$ refer to the states where two mutants are in distance of $1$ and $2$, respectively.

    \begin{figure}[!ht]
        \begin{adjustwidth}{-2.25in}{0in}
        \centering
        \includegraphics[width=\linewidth]{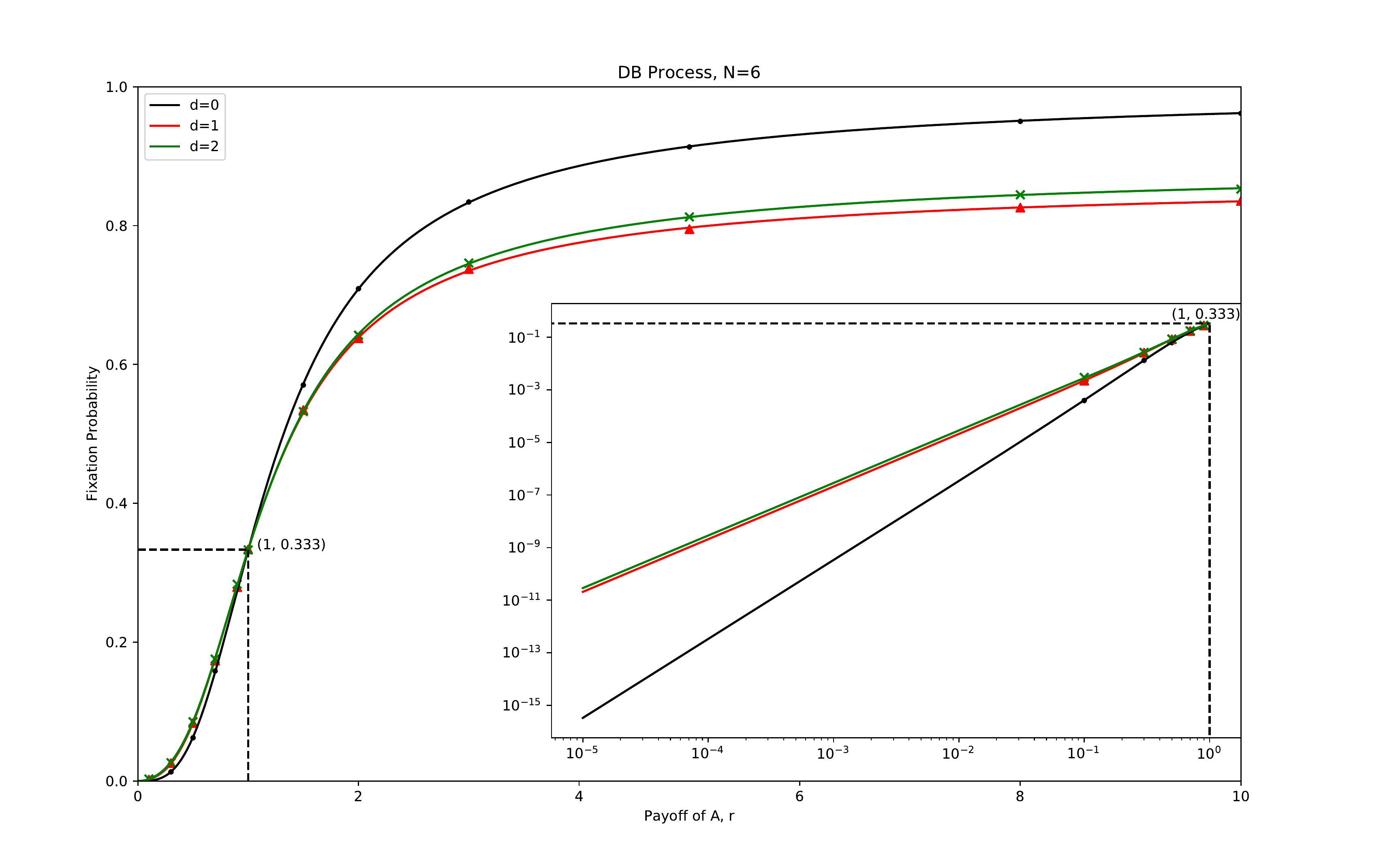}
        \caption{{\bf The fixation probability for $N=6$ with different mutant distances $d$.} The curve is drawn by the analytical results whereas the points are the simulation results. The iteration time for the simulation is $10^{6}$.
        }
        \label{Figure5}
        \end{adjustwidth}
    \end{figure}

    Base on Eq. \eqref{expand_6}, we find that on the circle with population size $6$:
    i) Under neutral selection, i.e. $r=1$, the fixation probabilities are only determined by the number of the mutants.
    It has nothing to do with the distance of mutants.
    ii) The first-order derivatives of fixation probabilities at $r=1$ for separated mutants are equal ($\frac{d}{dr}\Psi_{S_{1}} = \frac{d}{dr}\Psi_{S_{2}} = \frac{6}{12}$). They are greater than $0$, but are smaller than that of the connected mutants ($\frac{d}{dr}\Psi_{F_{2}} = \frac{7}{12}$). This indicates a slower change for the separated mutants than the connected mutants in fixation probability, as Fig. \ref{Figure5} shows. In particular, if $r<1$, i.e., the mutants are at a disadvantage, the fully connected mutants weaken the fixation probability. The fully connected mutants greatly promote the invasion when they are at an advantage (i.e., $r>1$).
    iii) The second-order derivative of the fixation probability at $r=1$ for $d=1$ is smaller than that for $d=2$. Here, the second-order derivative of the fixation probability is two times of the second-order coefficient of the Taylor series. 
    Thus, the closer the two mutants are, the less likely the invasion probability is under strong selection. 
    Consequecntly, the rank of the invasion chances is determined solely by the clustering factor, i.e., the distance of two mutants, as long as mutants are not fully connected.

\begin{figure}[!ht]
        \begin{adjustwidth}{-2.25in}{0in}
        \centering
        \includegraphics[width=\linewidth]{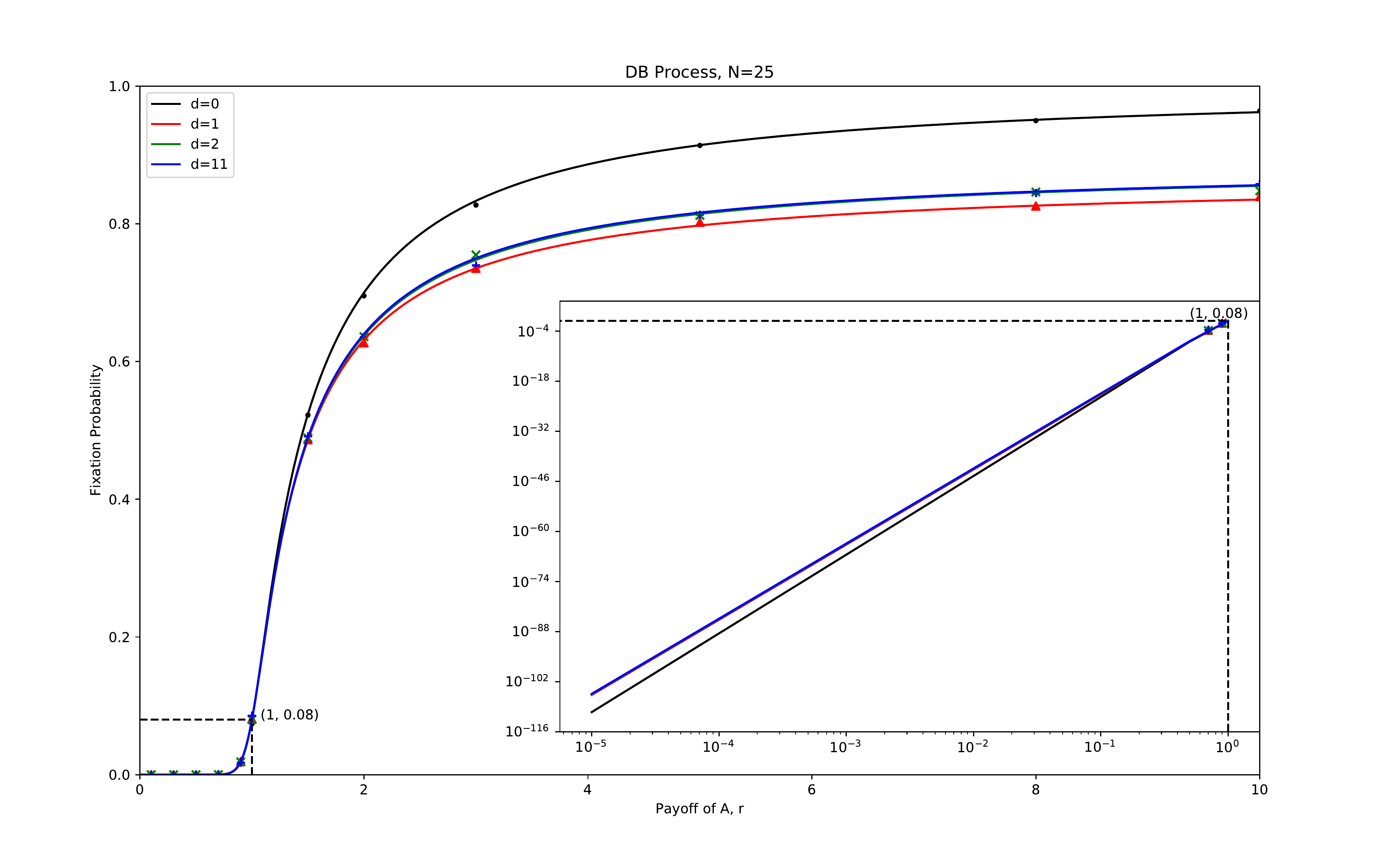}
        \caption{{\bf The fixation probability for $N=25$ with different mutant distances $d$.} 
        The curve is drawn by the results calculated by our developed algorithm whereas the points are the simulation results.
        It is noteworthy that the figure is quantitatively similar to Fig. \ref{Figure5}.
        }
        \label{Figure6}
        \end{adjustwidth}
    \end{figure}

    Using the developed algorithm, we generalize the above results on a small circle with population size $6$ to a circle with large size. We investigate the fixation probabilities for population size $25$ in Fig. \ref{Figure6}. Not all the distances are plotted with  only $d=1,2,3,11$ shown in Fig. \ref{Figure6}.
    This is because there are so many to show, and they do not lead to novel insights. 
    In addition, we list fixation probabilities and their derivatives at neutral selection numerically in Table \ref{derivatives} for population sizes $N=6, 25, 100$ respectively. We have found similar properties as that in the small one (Fig. \ref{Figure6}):
    i) The fixation probabilities are proportional to the number of mutants at neutral selection, i.e., $r = 1$.
    ii) The first-order derivatives at $r=1$ for separated mutants ($d > 0$) are the same.
    The first-order derivatives at $r=1$ for the connected mutants are greater than that of the separated mutants. 
    For $N=6,25,100$, we find that the first-order derivatives of fixation probabilities at neutral selection can be summarized as
    \begin{equation}
        \left\{
        \begin{aligned}
            & \frac{d}{dr}\Psi_{F_{2}}(=\Psi_{S_{0,1,1}}) \Bigg|_{r=1} & = & \frac{2N - 5}{2N} & \\
            & \frac{d}{dr}\Psi_{S_{d,1,1}} \Bigg|_{r=1} & = & \frac{2N - 6}{2N}&, d = 1, 2, ..., \lfloor\frac{N-2}{2}\rfloor
        \end{aligned}
        \right..
    \end{equation}
    This can apply $\forall N \geq 6$, but the proof is still an open issue.
    iii) For the separated mutants, the second-order derivative of the fixation probabilities at $r=1$ increases as the distance $d$ grows. Thus, the rank of the invasion chances is determined only by the assortment factor, provided that the mutants are not initially connected. In this case, the greater the distance two mutants are, the greater the fixation probabilities are. 
    Note that from $N=6$ to $N=25$, the second-order derivatives of the fixation probabilities at neutral selection increase from negative to positive.
    It implies that the fixation probability as a function of the selection intensity $r$ turns from convex to concave, as population size increases.
    \begin{table}[!ht]
        \begin{adjustwidth}{-2.25in}{0in}
        \centering
        \caption{
            The fixation probabilities of two mutants in distance $d$ and their derivatives at neutral selection, i.e., $r=1$.
            $N$ refers to the population size.}
        \begin{tabular}{|c||c|c|c||c|c|c||c|c|c|} \hline
        $N$ & \multicolumn{3}{c|}{$6$} & \multicolumn{3}{c|}{$25$} & \multicolumn{3}{c|}{$100$} \\\hline
        $d$ & $\Psi_{S_{d,1,1}}$ & $\frac{d}{dr}\Psi_{S_{d,1,1}}$ & $\frac{d^{2}}{dr^{2}}\Psi_{S_{d,1,1}}$ & $\Psi_{S_{d,1,1}}$ & $\frac{d}{dr}\Psi_{S_{d,1,1}}$ & $\frac{d^{2}}{dr^{2}}\Psi_{S_{d,1,1}}$ & $\Psi_{S_{d,1,1}}$ & $\frac{d}{dr}\Psi_{S_{d,1,1}}$ & $\frac{d^{2}}{dr^{2}}\Psi_{S_{d,1,1}}$ \\\hline
        $0$ & $0.333$ & $0.583$ & $-0.278$ & $0.08$ & $0.9$ & $5.1368$ & $0.02$ & $0.975$ & $29.9098$ \\\hline
        $1$ & $0.333$ & \textbf{$0.5$} & $-0.340$ & $0.08$ & \textbf{$0.88$} & $4.74230602$ & $0.02$ & \textbf{$0.97$} & $29.43617652$ \\\hline
        $2$ & $0.333$ & \textbf{$0.5$} & $-0.319$ & $0.08$ & \textbf{$0.88$} & $4.74983612$ & $0.02$ & \textbf{$0.97$} & $29.43805918$ \\\hline
        $3$ & - & - & - & $0.08$ & \textbf{$0.88$} & $4.75278453$ & $0.02$ & \textbf{$0.97$} & $29.4387966$ \\\hline
        ... & ... & ... & ... & ... & ... & ... & ... & ... & ...  \\\hline
        $10$ & - & - & - & $0.08$ & \textbf{$0.88$} & $4.75613829$ & $0.02$ & \textbf{$0.97$} & $29.43966386$ \\\hline
        $11$ & - & - & - & $0.08$ & \textbf{$0.88$} & $4.75617415$ & $0.02$ & \textbf{$0.97$} & $29.43968552$  \\\hline
        $12$ & - & - & - & - & - & - & $0.02$ & \textbf{$0.97$} & $29.4397024$ \\\hline
        ... & ... & ... & ... & ... & ... & ... & ... & ... & ...  \\\hline
        $48$ & - & - & - & - & - & - & $0.02$ & \textbf{$0.97$} & $29.43979022$ \\\hline
        $49$ & - & - & - & - & - & - & $0.02$ & \textbf{$0.97$} & $29.43979024$ \\\hline
        \end{tabular}
        \label{derivatives}
        \end{adjustwidth}
    \end{table}

    \subsection*{Conditional Fixation Times}
\begin{figure}[!ht]
        \begin{adjustwidth}{-2.25in}{0in}
        \centering
        \includegraphics[width=\linewidth]{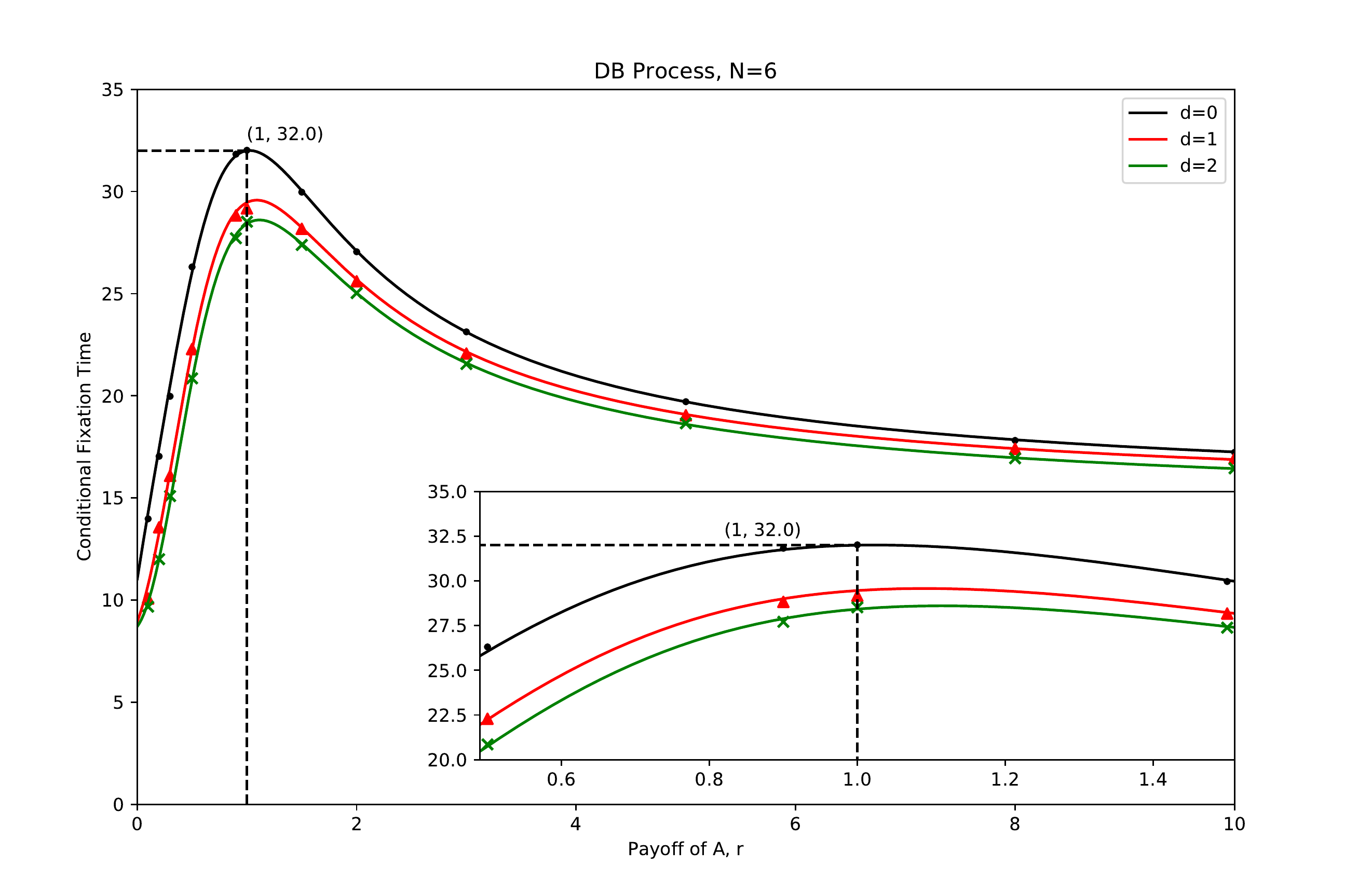}
        \caption{{\bf The conditional fixation time for two mutants in a circle of size $N=6$. The curve is the conditional fixation time obtained through Eq. \ref{ft_25_1} and the points are the simulation results. The iteration time for simulation is $10^5$. }}
        \label{Figure7}
        \end{adjustwidth}
    \end{figure}

    Taking $N=6$ into Eq. \eqref{ft_25_1}, we have the analytical results of the conditional fixation times $\tau_{S_{1}}^{A}$ and $\tau_{S_{2}}^{A}$. Due to the complexity of the expressions, we do not present them here.
    Expanding Eq. \eqref{ft_6_0}, $\tau_{S_{1}}^{A}$ and $\tau_{S_{2}}^{A}$ around neutral selection, i.e. $r=1$, results in
    \begin{equation}
        \left\{
        \begin{aligned}
            & \tau_{F_{2}}^{A} & = \frac{832}{26} &+& \frac{312}{416}(r-1) &-& \frac{78650}{4992}(r-1)^{2} + o((r-1)^{3}) \\
            & \tau_{S_{1}}^{A} & = \frac{767}{26} &+& \frac{1125}{416}(r-1) &-& \frac{85409}{4992}(r-1)^{2} + o((r-1)^{3}) \\
            & \tau_{S_{2}}^{A} & = \frac{739}{26} &+& \frac{1422}{416}(r-1) &-& \frac{88001}{4992}(r-1)^{2} + o((r-1)^{3})
        \end{aligned}
        \right. .
    \end{equation}
    Figure \ref{Figure7} presents the analytical predictions, which are validated by the simulations:
    i) At neutral selection $r=1$, the times for mutant fixation differ for different mutant distances, though the fixation probabilities are the same. The greater the distance two mutants is, the shorter it takes for the mutant to fixate. An intuitive explanation is that as the distance between two mutants grows, each mutant becomes more independent as a source for strategy spreading. This is similar to infection sources in epidemiology. More infection sources speed up mutant fixation. For instance, when two mutants are connected ($d=0$), there are initially only $2$ wild-type individuals that can be updated. On the contrary, when $d=2$, there are $4$ wild-type individuals that can be updated.
    ii) When the mutants are at an advantage ($r>1$), the conditional fixation time and the fixation probability are nontrivial. On the one hand, for each mutant distance, the mutant conditional fixation time grows at first and decrease as the mutant fitness $r$ grows. On the other hand, given a constant mutant fitness $r$, when the distance between two mutants $d$ becomes greater, the fixation time shrinks whereas the fixation probability changes non-monotonically (it decreases to the least when $d=1$ and grows when $d>1$).
    iii) When the mutants are disadvantageous ($r<1$), mutants have a better chance to fixate and also fixate faster as the distance between two mutants grows.
    In general, the fixation probability and the conditional fixation time for mutants do not have the same tendency as mutants are getting clustered \cite{phillip2017pre}. 
    And the rank of times for mutant fixation is monotonically determined by the clustering factor (i.e., the distance between mutants).

\begin{figure}[!ht]
        \begin{adjustwidth}{-2.25in}{0in}
        \centering
        \includegraphics[width=\linewidth]{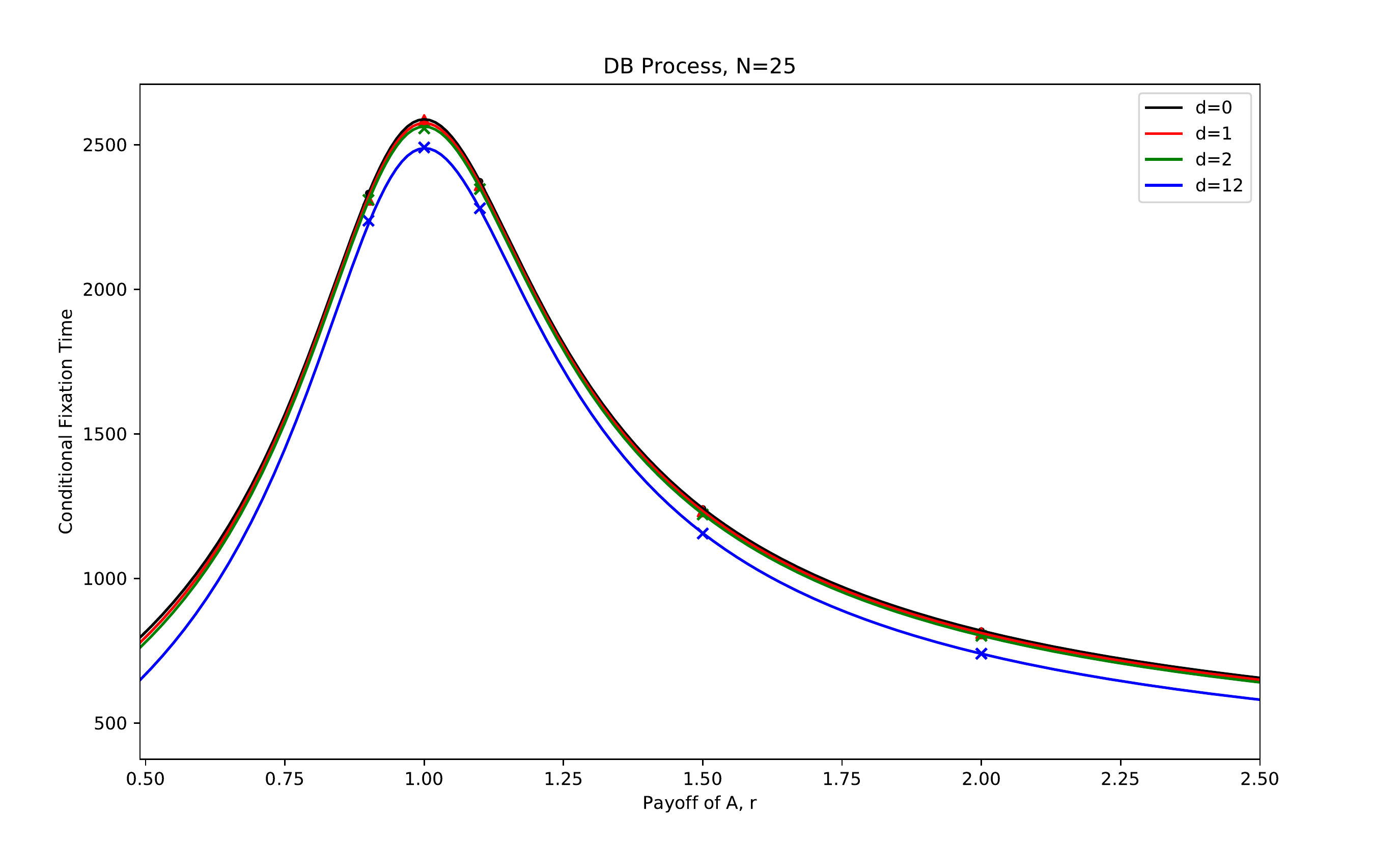}
        \caption{{\bf The conditional fixation time for mutants in a circle of size $N=25$ with different mutant distances $d$.} The curve is drawn by numerical results from our developed algorithm and points are obtained via simulation results. The iteration time for simulation is $10^6$.
        It is noteworthy that the figure is similar to Fig \ref{Figure7}, where the population size $N=6$. }
        \label{Figure8}
        \end{adjustwidth}
    \end{figure}

    Figure \ref{Figure8} presents the simulation results of the conditional fixation time of mutants and the numerical curve (calculated by our algorithm) for population size $N=25$.
    We observe that it agrees perfectly with the theoretical predictions.

\section*{Discussion}
Cooperation plays a key role in all levels of biological systems.
Network reciprocity, as one of the mechanisms to promote cooperation, has attracted considerable interests.
Network reciprocity results in an assortment between individuals using the same strategy \cite{assort1, assort2, assort2003, assort2012}.
Besides network reciprocity, the tag-based dynamics also yields more frequent interactions within groups equipping with the same tag \cite{rick2001nature, tag2, tag3}.
This in-group bias is created via the tag.
In this case, individuals who donate to others with the same tag
protect the cooperators from being exploited \cite{rick2001nature, tag2}.
The clustering individuals with the same tag lead to a high donation level.
Again, the clustering here shows itself as an important intermediate step to facilitate cooperation.
The assortment of mutants can either promote \cite{rule} or inhibit cooperation \cite{hauert2004nature}.
For example, the intensive interaction between individuals using the same strategy
can be beneficial for cooperation in the Prisoners' Dilemma,
yet can be destructive for cooperation in the Snowdrift Games \cite{taylor2006jtb}.
The nontrivial role the assortment plays can also be suggested from the previous studies: 
Fu et al. \cite{Fengfu2010} have compared invasions of the Snowdrift Game with that of the Prisoners' Dilemma on a lattice.
As the cost-to-benefit ratio grows, the mutants tend to emerge as few large compact clusters in the Prisoner's Dilemma whereas the mutants evolve to many dispersal small clusters in the Snowdrift Games.
All these previous studies are based on game interactions.
It is not clear how the assortment alone affects evolutionary dynamics.
Inspired by these,
we try to disentangle the spatial assortment of mutants from the game interaction.

We implemented a minimal model via adopting a circle as the spatial structure.
We assume that the fitness is frequency-independent.
And the assortment of two mutants is easily measured by the minimum number of wild-type individuals in between. 
In real biological systems, there can be three reasons for mutants to be spatially separated:
i) Independent mutations. The mutant individuals who are not spatially adjacent arise via independent mutations;
ii) Migration. One of the mutant individuals, originally adjacent to the others, migrate to another place and settle there; 
iii) The mutants are separated due to sudden environmental changes. 

As an illustrative case, we study the process with population size $N=6$.
It is the minimum size of a circle, where two mutants can be of three different distances (Fig. \ref{Figure2}).
We adopt the Death-birth process on a network.
The analytical results show that initially fully connected mutants enhance the group survival when they are at an advantage ($r>1$), whereas inhibit survival when mutants are at a disadvantage ($r<1$).
The simulation results are found to be in perfect agreement with the analytical ones (Fig. \ref{Figure5}).
In other words, the spatial assortment of mutants is an amplifier of natural selection for the connected mutants compared with the separated mutants.
However, as long as two mutants are separated,
the relative mutant fitness $r$ does not determine the rank of the probability of successful invasion, the distance between two mutants does.
Denoting $d$ as the initial distance between two mutants, the fixation probability falls to the smallest value when $d=1$, and it grows as $d$ becomes greater.
That is to say, as long as the mutants are separated, the further they initially are, the greater the invasion chance is.
It is true for both advantageous and disadvantageous mutants.
Our results show that the effect of spatial clustering on fixation is non-trivial,
even when the game interaction is absent.

The fixation probability for the separated mutants cannot be obtained as easily as obtaining that of the connected mutants.
On the one hand, the separated mutants introduce many additional states; 
on the other hand, the resulting Markov chain is not one-dimensional anymore.
We further categorize the transient states into two classes.
And we make use of the fixation probabilities for the connected mutants to obtain the fixation probabilities for the separated mutants.
In addition, we have developed an efficient algorithm to estimate the fixation probability for the separated mutants in arbitrary population size.
In general, the algorithm consists of three steps:
1) Listing all the states of the Markov chain in order,
2) Listing all the transition probabilities in \nameref{S2_Appendix},
3) calculate the derivatives in Eq. \eqref{1st_derivative} and Eq. \eqref{2nd_derivative} based on \nameref{S3_Appendix}.
The time complexity is of $O(N^{4.746})$. 
The space complexity is of $O(N^{4})$, and $O(N^{2})$
if sparse matrix methods are adopted \cite{algo_complex, algo_complex2, algo_complex3}.
We evaluated the processes for population sizes of $25$ and $100$.
All the above-mentioned results still apply as in the small circle with size $6$.
Thus, we conjecture that the main results apply to any population size.
Yet the strict proof is still an open issue.

In the work of Ohtsuki et al. \cite{rule}, the pair approximation is adopted.
Under weak selection limit, 
it is only the initial frequency of mutants that is key to the fixation probabilities.
The fixation probability has nothing to do with the assortment parameter.
There, the assortment parameter is given by $q_{A|A} - q_{B|A}$ \cite{assort2003, indexassort2013, indexassort2016},
which refers to the difference between the number of neighbors using the same strategy and that of the ones using the other strategy.
Our results show a different picture.
Let us take a ring consisting of $25$ individuals with $2$ mutants as an example.
When the distance between two mutants are equal or greater than two ($2 \leq d \leq 12$),
all the cases share the same probability of finding a mutant next to a wild-type individual ($q_{B|A}=\frac{4}{23}$) or finding a wild-type individual next to a wild-type individual ($q_{A|A}=1$).
By pair approximation, the fixation probability is the same for all the mentioned initial population configurations,
since the number of mutants is two, and even the assortment factor $q_{A|A}-q_{B|A}=\tfrac{19}{23}$ is the same.
However, our analytical results show that the fixation probabilities can be in a large difference when the selection intensity is strong, verified by simulations.
Furthermore, we show that the difference occurs at the second-order derivative of the selection intensity.
This suggests that the number of initial mutants alone cannot determine the fixation probability.
Therefore, the pair approximation is not sufficient to portray the spatial clustering accurately,
provided the selection intensity is strong.

We investigate the cases with Death-birth process. 
The DB process contains two steps: 
An individual is randomly chosen to die and its nearest neighbors compete to reproduce an identical offspring. 
The assumption behind the DB process is that the death rate is equal for all the individuals and the selection happens in the stage of reproduction. The competition for reproduction is local.
The Birth-death (BD) process is different from the DB process.
It contains two steps as well:
An individual is chosen to produce an identical offspring with the possibility proportional to its fitness across the entire population, 
and the offspring replaces a neighbor of its parent randomly.
The assumption here is that all the individuals compete for reproduction, thus the selection is global. The death rate is equal for all the neighbors. 
Noteworthily, for the BD process,
the isothermal theorem \cite{nature_05} shows that the fixation probability of mutants with BD process is identical to that in well-mixed populations on all the isothermal graphs.
A circle is an isothermal graph.
It implies that the assortment of the mutants does not play any role in fixation probability,
provided the number of mutants is the same for a circle with BD process.
However, our results with the DB process depicts a different picture.
Consequently, the details of update rule could dramatically alter the evolutionary dynamics on networks,
even if the game interaction is not at work.
Intuitively, the assortment plays a role if the competition for reproduction is local, but is not at work if the competition for reproduction is global.
In fact, it has been shown that different evolutionary rules can alter the evolutionary outcome not only in the networked population \cite{rule} but also in a simple well-mixed population \cite{wu15njp}.
Our results also echo the recent studies that the DB process does not conform to the isothermal theorem \cite{kamran2015royal,laura2015plos}.

To sum up, our results reveal counterintuitive but fundamental effects of spatial clustering on the evolutionary dynamics.
In particular, the clustering plays its role without the involvement of games.
It is not hard to imagine the great complexity arises when games are involved or when complex graphs are introduced.
This deserves further studies.
In addition,
our model can be used as a reference case to better understand how the clustering of mutants favor or disfavor cooperation.
Furthermore, it is a natural association that our conclusion calls for biological experiments, such as the microbial experiments, to verify the effect of spatial assortment on evolutionary dynamics.


\section*{Supporting information}

\paragraph*{S1 File.}
\label{S1_File}
{\bf Python program.} We implemented our developed algorithm in a Python program to calculate the theoretical fixation probabilities and the derivatives. The code is also available at \url{https://github.com/Awdrtgg/DB_on_cycle/blob/master/fp_and_derivatives.py}.

\paragraph*{S2 File.}
\label{S2_File}
{\bf Python program.} We implemented our developed algorithm in a Python program to calculate the theoretical conditional fixation times. The code is also available at \url{https://github.com/Awdrtgg/DB_on_cycle/blob/master/fixation_time.py}.

\paragraph*{S3 File.}
\label{S3_File}
{\bf Python program.} We evaluated the fixation probabilities and fixation times by simulating the process in a Python program. The simulation code is also available at \url{https://github.com/Awdrtgg/DB_on_cycle/blob/master/simulation.py}.

\paragraph*{S1 Appendix.}
\label{S1_Appendix}
{\bf Algorithms for transformation between the state number and the triplet.}

\paragraph*{S2 Appendix.}
\label{S2_Appendix}
{\bf The algorithm to obtain the transition matrices.}

\paragraph*{S3 Appendix.}
\label{S3_Appendix}
{\bf The first-order and second-order derivatives of fixation probability.}


    \newpage
    {\bf S1 Appendix. Algorithms for transformation between the state number and the triplet.} 
    
    Functions $triplet\_to\_state$ and $state\_to\_triplet$ shows the transformations between the triplet and the state.

    \begin{algorithm}
        \caption{triplet\_to\_state($n, x, a, b$) to map the triplet to the state}
        \label{state_to_num}
        \renewcommand{\algorithmicrequire}{\textbf{Input:}}
        \renewcommand{\algorithmicensure}{\textbf{Output:}}
        \begin{algorithmic}[1]
            \REQUIRE
            \ \\
            The total individual number $N$; \\
            The triplet $x, a, b$; \\
            \ENSURE The state
            \IF {$a > b$}
                \STATE swap($a, b$) // Relation constraint
            \ENDIF

            \IF {$x > n - a - b$}
                \STATE $x \leftarrow n - a - b$ // Relation constraint
            \ENDIF

            \IF {$x = 0$}
                \RETURN $F_{a+b}$
            \ENDIF

            \IF {$a = 0$}
                \RETURN $F_{b}$
            \ENDIF

            \STATE $i \leftarrow 0$
            \FOR {$w \leftarrow 0$ \TO $(a + b)$}
                \STATE $i \leftarrow i + \lfloor\frac{w}{2}\rfloor \cdot \lfloor\frac{n-w}{2}\rfloor$
            \ENDFOR
            \STATE $i \leftarrow i + (a-1) \cdot \lfloor\frac{n-a-b}{2}\rfloor$
            \STATE $i \leftarrow i + x - 1$
            \RETURN $S_{i}$
        \end{algorithmic}
    \end{algorithm}

    \begin{algorithm}
        \caption{state\_to\_triplet($i$) to map the state number to the triplet}
        \label{num_to_state}
        \renewcommand{\algorithmicrequire}{\textbf{Input:}}
        \renewcommand{\algorithmicensure}{\textbf{Output:}}
        \begin{algorithmic}[1]
            \REQUIRE
            \ \\
            The state number $i$; \\
            \ENSURE $x, a, b$
            \STATE $a\leftarrow 1$
            \STATE $w \leftarrow 0$
            \WHILE {$i \geq \lfloor\frac{w}{2}\rfloor \cdot \lfloor\frac{n-w}{2}\rfloor$}
                \STATE $i \leftarrow i - \lfloor\frac{w}{2}\rfloor \cdot \lfloor\frac{n-w}{2}\rfloor$
                \STATE $w \leftarrow w + 1$ // To obtain the number of mutants
            \ENDWHILE

            \WHILE {$i \geq \lfloor\frac{n-w}{2}\rfloor$}
                \STATE $i \leftarrow i - \lfloor\frac{n-w}{2}\rfloor$
                \STATE $a \leftarrow a + 1$ 
            \ENDWHILE

            \STATE $b \leftarrow w - a$
            \STATE $x \leftarrow i + 1$
            \RETURN $x, a, b$
        \end{algorithmic}
    \end{algorithm}

    \newpage
    {\bf S2 Appendix. The algorithm to obtain the transition matrices.} 
    
    Algorithm \ref{main_algo} shows how to obtain the transition matrices $Q_{1}$ and $Q_{2}$. There are at most $8$ operations for each state. As there are totally $O(N^{2})$ states, the time complexity of the algorithm is $O(N^{2})$. And the space complexity of the algorithm are the same as the scale of the matrix $Q_{1}$, which is $O(N^{4})$. It is noteworthy that $Q_{1}$ and $Q_{2}$ are sparse matrices when $N$ is great. Therefore, the space complexity is $O(N^{2})$ if sparse storing methods are adopted.

    \begin{algorithm}[H]
        \caption{Calculation of $Q_{1}$ and $Q_{2}$}
        \label{main_algo}
        \renewcommand{\algorithmicrequire}{\textbf{Input:}}
        \renewcommand{\algorithmicensure}{\textbf{Output:}}
        \begin{algorithmic}[1]
            \REQUIRE
            \ \\
            Total individual number $N$; \\
            Mutant payoff $r$; \\
            \ENSURE $Q_{1}, Q_{2}$
            \STATE $Total \leftarrow 0$ // The number of the states

            \FOR {$w \leftarrow 0$ \TO $N+1$}
                \STATE $Total \leftarrow Total + \lfloor\frac{w}{2}\rfloor \cdot \lfloor\frac{n-w}{2}\rfloor$ 
            \ENDFOR

            \STATE $Q_{1} \leftarrow [0]_{Total\times Total}$
            \STATE $Q_{2} \leftarrow [0]_{Total\times (N+1)}$

            \FOR {$i \leftarrow 1$ \TO $Total$}
                \STATE $prob\_self \leftarrow 1$ // The probability not to transit to another state
                \FOR {$j \leftarrow 1$ \TO $11$}
                    \STATE $x, a, b \leftarrow state\_to\_triplet( i)$

                    \STATE Change $x, a, b$ according to Table 2

                    \STATE $t \leftarrow triplet\_to\_state(N, x, a, b)$
                    \STATE Assign $temp\_prob$ according to Table 2
                    \IF {$t \in F$}
                        \STATE $Q_{2}[S_{i}, t] \leftarrow Q_{2}[S_{i}, t] + temp\_prob$
                    \ELSE
                        \STATE $Q_{1}[S_{i}, t] \leftarrow Q_{1}[S_{i}, t] + temp\_prob$
                    \ENDIF
                    \STATE $prob\_self \leftarrow prob\_self - temp\_prob$
                \ENDFOR
                \STATE $Q_{1}[S_{i}, S_{i}] \leftarrow Q_{1}[S_{i}, S_{i}] + prob\_self$
            \ENDFOR

            \RETURN $Q_{1}, Q_{2}$
        \end{algorithmic}
    \end{algorithm}

    \newpage
    {\bf S3 Appendix. The first-order and second-order derivatives of fixation probability.} 

    In the following, we give the details of calculating the first-order derivatives and the second-order derivatives of the fixation probability $\Psi$. In general, we have
    \begin{equation}
        \Psi(r) = [I - Q_{1}(r)]^{-1} Q_{2} \pi(r).
    \end{equation}
    Where $I$ is an identity matrix of the same size with $Q_{1}$. The derivation of $\Psi$ is
    \begin{equation}
        \label{calc_deri_psi}
        \frac{d}{dr} \Psi(r) = \frac{d}{dr}[I - Q_{1}(r)]^{-1} Q_{2} \pi(r) + [I - Q_{1}(r)]^{-1} Q_{2} \frac{d}{dr}\pi(r).
    \end{equation}
    Since we have the analytical result for $\pi(r)$, the second part of the right side of equation is acknowledged. The point of the problem comes to $\frac{d}{dr}[I - Q_{1}(r)]^{-1}$.

    The property of the matrix inversion gives rise to
    \begin{equation}
        [I - Q_{1}(r)]^{-1}[I - Q_{1}(r)] = I.
    \end{equation}
    The derivations of both sides of the equation is
    \begin{equation}
        \frac{d}{dr}[I - Q_{1}(r)]^{-1} [I - Q_{1}(r)] + [I - Q_{1}(r)]^{-1} \frac{d}{dr}[I - Q_{1}(r)] = 0.
    \end{equation}
    This is equivalence to
    \begin{equation}
        \frac{d}{dr}[I - Q_{1}(r)]^{-1} [I - Q_{1}(r)] = [I - Q_{1}(r)]^{-1} \frac{d}{dr} Q_{1}(r).
    \end{equation}
    Multiply $[I - Q_{1}(r)]^{-1}$ for both sides, we have
    \begin{equation}
        \label{calc_deriv_i-q1}
        \frac{d}{dr}[I - Q_{1}(r)]^{-1} = [I - Q_{1}(r)]^{-1} \frac{d}{dr}Q_{1}(r)[I - Q_{1}(r)]^{-1}.
    \end{equation}
    Taking Eq. \eqref{calc_deriv_i-q1} into Eq. \eqref{calc_deri_psi} results in
    \begin{equation}
        \frac{d}{d r}\Psi(r) = [I - Q_{1}(r)]^{-1}\frac{d}{d r}Q_{1}(r)[I - Q_{1}(r)]^{-1}Q_{2}\pi(r) + [I - Q_{1}(r)]^{-1}Q_{2}\frac{d}{d r}\pi(r),
    \end{equation}
    And the second-order derivative can be obtained similarly.


\begin{thebibliography}{15}

    \bibitem{Hamilton1964jtb}
    Hamilton WD.
    The genetical evolution of social behaviour. I. Journal of Theoretical Biology. 1964 Jul; 7(1):1-16.

    \bibitem{rick2001nature}
    Riolo RL, Cohen MD, Axelrod R. Evolution of cooperation without reciprocity. Nature. 2001 Nov;414(6862):441.

    \bibitem{lega2003evolution}
    Le Galliard JF, Ferrière R, Dieckmann U. The adaptive dynamics of altruism in spatially heterogeneous populations. Evolution. 2003 Jan;57(1):1-7.

    \bibitem{hauert2004nature}
    Hauert C, Doebeli M. Spatial structure often inhibits the evolution of cooperation in the snowdrift game. Nature. 2004 Apr;428(6983):643.

    \bibitem{five_rules}
    Nowak MA. Five rules for the evolution of cooperation. science. 2006 Dec 8;314(5805):1560-3.

    \bibitem{proc_06_circle}
    Ohtsuki H, Nowak MA. Evolutionary games on circles. Proceedings of the Royal Society of London B: Biological Sciences. 2006 Sep 7;273(1598):2249-56.

    \bibitem{langer2008jtb}
    Langer P, Nowak MA, Hauert C. Spatial invasion of cooperation. Journal of Theoretical Biology. 2008 Feb 21;250(4):634-41.

    \bibitem{hauert2009}
    Traulsen A, Hauert C. Stochastic evolutionary game dynamics. Reviews of nonlinear dynamics and complexity. 2009 Jul 10;2:25-61.

    \bibitem{assort2012}
    Van Veelen M, García J, Rand DG, Nowak MA. Direct reciprocity in structured populations. Proceedings of the National Academy of Sciences. 2012 Jun 19;109(25):9929-34.

    \bibitem{kamran2015royal}
    Kaveh K, Komarova NL, Kohandel M. The duality of spatial death–birth and birth–death processes and limitations of the isothermal theorem. Royal Society open science. 2015 Apr 1;2(4):140465.

    %

    \bibitem{rule}
    Ohtsuki H, Hauert C, Lieberman E, Nowak MA. A simple rule for the evolution of cooperation on graphs and social networks. Nature. 2006 May;441(7092):502.

    %

    \bibitem{wu2016plos}
    Jorge Peña, Wu B, Arranz J, Traulsen A. Evolutionary games of multiplayer cooperation on graphs. PLoS Computational Biology, 2016 Jul;12(8), e1005059.

    %

    \bibitem{assort1}
    Toro M Silio, L. Assortment of encounters in the two-strategy game. Journal of theoretical biology, 1986; 123(2), 193-204.

    \bibitem{assort2}
    Wilson DS, Dugatkin LA. Group selection and assortative interactions. The American Naturalist. 1997 Feb 1;149(2):336-51.

    %

    \bibitem{wu2015pre}
    Sui X, Wu B, Wang L. Speed of evolution on graphs. Physical Review E, 2015, 92(6):062124.

    %

    \bibitem{tag3}
    Wu T, Wang L, Fu F. Coevolutionary dynamics of phenotypic diversity and contingent cooperation. PLoS computational biology. 2017 Jan 31;13(1):e1005363.

    \bibitem{tag2}
    Traulsen A, Schuster HG. Minimal model for tag-based cooperation. Physical Review E. 2003 Oct 27;68(4):046129.

    %

    \bibitem{nowak2006book}
    Nowak MA. Evolutionary dynamics. Harvard University Press. 2006 Sep 29.

    %

    \bibitem{bin2018aspiration}
    Wu B, Zhou L. Individualised aspiration dynamics: Calculation by proofs. PLoS computational biology. 2018 Sep 25;14(9):e1006035.

    %

    \bibitem{firstcourse1975}
    Karlin S, Taylor HM. A first course in stochastic processes. Academic Press. 1975.

    %

    \bibitem{hinderson2014counter}
    Hindersin L, Traulsen A. Counterintuitive properties of the fixation time in network-structured populations. Journal of The Royal Society Interface. 2014 Oct 6;11(99):20140606.

    %

    \bibitem{grimmett2009book}
    Grimmett G, Stirzaker D. Probability and random processes. Oxford university press; 2001 May 31 (pp. 221-224).

    %

    \bibitem{algo_complex}
    Coppersmith D, Winograd S. Matrix multiplication via arithmetic progressions. Journal of symbolic computation. 1990 Mar 1;9(3):251-80.

    \bibitem{algo_complex2}
    Le Gall F. Powers of tensors and fast matrix multiplication. InProceedings of the 39th international symposium on symbolic and algebraic computation. 2014 Jul 23 (pp. 296-303). ACM.

    \bibitem{algo_complex3}
    Hindersin L, Wu B, Traulsen A, García J. Computation and simulation of evolutionary Game Dynamics in Finite populations. Scientific Reports. 2019 May 6;9(1):6946.

    %
    
    \bibitem{phillip2017pre}
    Altrock, PM, Traulsen A, Nowak MA. Evolutionary games on cycles with strong selection. Physical Review E. 2017;95:022407

    %

    \bibitem{assort2003}
    Bergstrom TC. The algebra of assortative encounters and the evolution of cooperation. International Game Theory Review. 2003 Sep;5(03):211-28.

    %

    \bibitem{taylor2006jtb}
    Taylor C, Nowak MA. Evolutionary game dynamics with non-uniform interaction rates. Theoretical population biology. 2006 May 1;69(3):243-52.

    %

    \bibitem{Fengfu2010}
    Fu F, Nowak MA, Hauert C. Invasion and expansion of cooperators in lattice populations: Prisoner's dilemma vs. snowdrift games. Journal of theoretical biology. 2010 Oct 7;266(3):358-66.

    %

    \bibitem{indexassort2013}
    Alger I, Weibull JW. A generalization of Hamilton's rule—Love others how much?. Journal of Theoretical Biology. 2012 Apr 21;299:42-54.

    \bibitem{indexassort2016}
    Nax HH, Rigos A. Assortativity evolving from social dilemmas. Journal of theoretical biology. 2016 Apr 21;395:194-203.

    %

    \bibitem{nature_05}
    Lieberman E, Hauert C, Nowak MA. Evolutionary dynamics on graphs. Nature. 2005 Jan;433(7023):312.

    %

    \bibitem{wu15njp}
    Wu B, Bauer B, Galla T, Traulsen A. Fitness-based models and pairwise comparison models of evolutionary games are typically different—even in unstructured populations. New Journal of Physics. 2015 Feb 13;17(2):023043.

    %

    \bibitem{laura2015plos}
    Hindersin L, Traulsen A. Most undirected random graphs are amplifiers of selection for birth-death dynamics, but suppressors of selection for death-birth dynamics. PLoS computational biology. 2015 Nov 6;11(11):e1004437.

    %











    \end{thebibliography}
    \end{document}